\documentstyle[epsf,aps,12pt]{revtex} 
\begin{document} 
\pagestyle{plain} 
\setcounter{page}{1} 
\baselineskip=0.3in 
\begin{titlepage} 
\vspace{.5cm} 
 
\begin{center} 
{\Large Top decays into lighter stop and gluino   } 

\vspace{0.2in}
Lian You Shan $^{a,b}$ and Shou Hua Zhu $^{a,b}$  \\
$a$ CCAST (World Lab), P.O. Box 8730, Beijing 100080, P.R. China \\
$b$ Institute of Theoretical Physics, Academia Sinica, \\
P.O. Box 2735, Beijing 100080, P.R. China 

\end{center} 

\begin{footnotesize} 
\begin{center}\begin{minipage}{5in} 
\baselineskip=0.25in 
\begin{center} ABSTRACT\end{center}  
We have calculated the decay width of the process $t \rightarrow
\tilde{g} \tilde{t}_1$ including one loop QCD corrections.
We found that the decay width of the such process could be comparable 
 with that of the standard channel $t \rightarrow b W^+$, 
 and, its QCD correction could enhance the widths over $30 \%$
  in a very large mass range of the lighter stop. 

\end{minipage} 
\end{center}
\end{footnotesize} 

\vfill

\end{titlepage} 
 
\newpage 

\section{Introduction} 

Top quark play an important role in the search for new physics due to its 
huge mass compared to other fermions in the standard model (SM).
Besides its large Yukawa coupling to Higgs sector, top quark also 
anticipates strong interaction (QCD) with its supersymmetric partner through 
the top-stop-gluino vertex in a particle model of supersymmetry (SUSY).
As a supersymmetric counterpart, a large mass splitting can emerge between 
the mass-eigenstates of the stops and could lead to a lighter physical stop, 
so that stop quark may be produced in our top quark factories.

The experimental lower mass limit of stop is 86 GeV \cite{lep2} and the 
theoretical one could be more smaller \cite{Ellis}.
In most of conventional proposals stop will be produced in pair, 
then they need larger energy input in general.
 Comparing to stop pair production, 
one wonder whether stop could be produced
 associated with a light gluino in the decay of top quark. 
Such a channel can be accommodated in the single top events, 
and the top quark sector would become a nice place to study SUSY-QCD.  

The key point lies in that, whether the gaugino of $SU(3)_c$ is also 
sufficiently light. 
We would like to make an analyses on the situation of light gluino bellow.

Historically, there had ever glimmered light gluinos in many scenarios with 
natural assumptions about the mechanics of SUSY breaking at higher energy scale.
For example, the mass of gluino was resulted in $1 \sim 100  $ GeV typically if the SUSY breaking was transmitted to the observable sector by gravity (mSUDRA); 
 and the masses of gluinos and squarks will arise smaller ( bellow GeV ) as a loop correction (next leading order effect) when that breaking is gauge-mediated ( GMSB ) \cite{farra}. 
Those light gluino scenario can explain several well-established phenomena,
such as the anomalously slow running of $SU(3)_c$ coupling,
anomalies in jet production \cite{jets} as well as in 
the $\eta(1410) \cite{eta}$.

The difficulty is obviously that, present colliders do not favor light gluino.
A lower mass bound $180 $ GeV has ever been published \cite{liex}. 
Nevertheless, a controversial assumption that there is no gluino lighter 
than $5$ GeV as well as the state-of-art usage of perturbative theory 
\cite{alep}, had made the constraint less convincing,
especially leave a margin for a gluino light than $5$ GeV. 
The concomitant predictions of a light Higgs mass and a light chargino,
have also been (at least marginally) ruled out at LEP II\cite{Opal97}, 
so has been the light gluino in mSUDRA models. 
However, light gluino in GMSB models can 
survive in such experiment.
Anyway all the direct searches for a light gluino have turned to 
a negative results \cite{nega} since 1998. 

Then light gluino is cornered into bound states (for example $R^0$ ) 
to live if it is light enough.
For a light gluino in a bound state, the well known KTeV experiment is an 
rather important one to give a constraint on its mass. KTeV result indeed 
has closed most of the place for a light gluino \cite{KTeV} and has given 
an impression that light gluino had been ruled out. Although, there are still 
two shreds that are not excluded by KTeV data in fact.
The gluino could weight as $ 0 \sim 1.6 GeV $ or as $ 5 GeV $ \cite{KTeV}.
Such light gluino can survive through another important E761 experiment too
\cite{E761}. 

Although recent re-analyses\cite{slow} showed a consistency with SM 
and really eroded the attractiveness of light gluino in explaining
 slow running of $SU(3)_c$ coupling, such re-analyses relied too much 
on the measurement of low energy QCD, for example $\alpha_s(M_{\tau})$, 
which itself is difficult enough. 
 
Through above investigation, we have make it clear that, 
there is no counter-indication which can conclusively convince one
to give up a light gluino ( few GeV or so).
Recently, more new suggestion on light gluino within the framework of GMSB
have been proposed\cite{more}. Both scenario allow a gluino at few GeV and 
predicted the standard missing energy signature of SUSY 
may become not suitable.

Then from the narrow window for light gluino, a noticeable partial width to 
stop and gluino might be detected in top quark decay when the statistics 
of the top events are improved.

In the present paper, we  
will consider the decay process $t \rightarrow \tilde{g}\tilde{t}_1$ 
including the QCD corrections, which possibility has been mentioned 
by Raby et. al\cite{more}.
In experiments, 
the stop may be followed by the decays 
$\tilde{t}_1 \rightarrow c {\tilde \chi}^0_1$ or 
$\tilde{t}_1 \rightarrow b {\tilde \chi}^+_1$ \cite{stoglude}, then its typical 
signal may be one hadronic jet plusing a large missing $P_T$.  
Instead of the detailed analysing the signal against possible background, 
we emphasize the possible anomalous branch ratio of top decay. 
In fact, in the SM  $BR(t \rightarrow b W^+) $ had 
been assumed as the dominant one for the observed top pair events at 
Tevatron \cite{exp}.
 Thus, the non-SM branching ratio of the top
 quark which could be as large as the SM one is not excluded, 
 and the searching 
 for non-SM top decay are undergoing in both CDF and $D\O$ \cite{nonsm}. 

In the literatures, there have been several works \cite{Beenakker}
discussing the
decays of stop or gluino into top quark, 
which have the same dynamics (interaction vertex) as the process
discussed here, 
and these calculations offered a good references to check our calculations. 

\section{Calculations} 

The tree level diagram of the process $t(p) \rightarrow \tilde{g} (k_1) 
\tilde{t}_1 (k_2)$ is  shown  in Fig. 1  (a), and the decay width is
\begin{eqnarray}
\Gamma_0={N \alpha_s \over 3 m_t^3} (m_t^2-m_{\tilde{g}}^2+m_{\tilde{t}_1}^2-
2 m_{\tilde{g}} m_{\tilde{t}_1} \sin(2\theta)), 
\end{eqnarray}
where
\begin{eqnarray}
N=\sqrt{(m_t^2-(m_{\tilde{g}}+m_{\tilde{t}_1})^2)
(m_t^2-(m_{\tilde{g}}-m_{\tilde{t}_1})^2)},
\end{eqnarray}
and $\theta$ is the stop mixing angle which is to be defined in next section.

\subsection{Virtual corrections}

The $O(\alpha_s)$ virtual corrections arise from 
triangle as well as 
the gluino, quark and squark 
self energies diagrams as shown in Fig. 1 ($b \sim e$) respectively. Through our 
calculations, 
we employed dimensional reduction regularization (DRR) to control all the 
ultraviolet divergences, which is widely adopted in the
calculations of the radiative corrections in the minimal supersymmetric
standard model (MSSM) since the conventional dimensional regularization 
violates supersymmetry with a mismatch between the numbers of degree of 
freedom of gauge bosons and gauginos\cite{drr}.
At the same time,
we adopted the on-mass-shell renormalization scheme \cite{r13,r14}.  

The one loop decay width can be expressed as
\begin{eqnarray}
\Gamma_1&=& {4 \sqrt{2} \pi N \alpha_s^2 \over 3 m_t^3} Re
\left( f_1((m_{\tilde{g}}+m_t)^2-m_{\tilde{t}_1}^2) (\cos\theta - \sin\theta) \right.
\nonumber \\
&& -
\left.
f_2((m_{\tilde{g}}-m_t)^2-m_{\tilde{t}_1}^2)(\sin\theta+\cos\theta) \right),
\end{eqnarray}
where
\begin{equation}
f_1=f_1^c+f_1^d, \ \ \ \ f_2=f_2^c+f_2^d,
\label{total1}
\end{equation}
where $f^c_i$ are the contributions in the form of counterterms, 
\begin{eqnarray}
f_1^c&=&\sqrt{2}(f_L\cos\theta-f_R \sin\theta), \nonumber \\
f_2^c&=&\sqrt{2}(-f_L\cos\theta-f_R \sin\theta), \nonumber \\
f_L&= &{1\over 2} Z_L^t+ {1\over 2} Z_R^{\tilde{g}} +\delta g_s/g_s -
\tan\theta \delta\theta+ {1\over 2} Z_{\tilde{t}_1}, \nonumber \\
f_R&= &{1\over 2} Z_R^t+ {1\over 2} Z_L^{\tilde{g}} +\delta g_s/g_s +
\cot\theta \delta\theta+ {1\over 2} Z_{\tilde{t}_1}.
\label{total}
\end{eqnarray}
The renormalization constants in Eq. \ref{total} are,
\begin{eqnarray}
Z_L^t&=& {1 \over 12 \pi^2 m_t^2}
\left( 
\sin\theta^2 (m_{\tilde{t}_1}^2-m_{\tilde{g}}^2) B_0(0,m_{\tilde{g}}^2,m_{\tilde{t}_1}^2)
+\cos\theta^2 (m_{\tilde{t}_2}^2-m_{\tilde{g}}^2) B_0(0,m_{\tilde{g}}^2,m_{\tilde{t}_2}^2)
\right. \nonumber \\
&&-m_t^2 B_0(0,\lambda^2,m_t^2) +
\left( \sin\theta^2 (m_{\tilde{g}}^2-m_{\tilde{t}_1}^2)-\cos\theta^2 m_t^2 \right)
B_0(m_t^2,m_{\tilde{g}}^2,m_{\tilde{t}_1}^2) \nonumber \\
&& +\left( \cos\theta^2 (m_{\tilde{g}}^2-m_{\tilde{t}_2}^2)-\sin\theta^2 m_t^2 \right)
B_0(m_t^2,m_{\tilde{g}}^2,m_{\tilde{t}_2}^2)  \nonumber \\
&&+
m_t^2 (-m_{\tilde{g}}^2+m_{\tilde{t}_1}^2-m_t^2 +
2 m_{\tilde{g}} m_t \sin(2 \theta)) \dot{B}(m_t^2,m_{\tilde{g}}^2,m_{\tilde{t}_1}^2)
\nonumber \\
&& +m_t^2 (-m_{\tilde{g}}^2+m_{\tilde{t}_2}^2-m_t^2 -
2 m_{\tilde{g}} m_t \sin(2 \theta)) \dot{B}(m_t^2,m_{\tilde{g}}^2,m_{\tilde{t}_2}^2)
\nonumber \\
&& 
\left.+4 m_t^4 \dot{B}(m_t^2, \lambda^2,m_t^2) \right), \nonumber \\
Z_R^t&=& Z_L^t (\theta \rightarrow \pi/2-\theta), \nonumber \\
Z_L^{\tilde{g}}&=&{1 \over 32 \pi^2 m_{\tilde{g}}^2}
\left( 
\sum_{\tilde{q}}\{ m_{\tilde{q}_L}^2 B_0(0, m_q^2,m_{\tilde{q}_L}^2)
+ m_{\tilde{q}_R}^2 B_0(0, m_q^2,m_{\tilde{q}_R}^2) \right. 
\nonumber \\
&&-(m_{\tilde{g}}^2+m_{\tilde{q}_L}^2) B_0(m_{\tilde{g}}^2, m_q^2,m_{\tilde{q}_L}^2)
-(m_{\tilde{g}}^2+m_{\tilde{q}_R}^2) B_0(m_{\tilde{g}}^2, m_q^2,m_{\tilde{q}_R}^2)
\nonumber \\
&& +2 m_{\tilde{g}}^2 (m_{\tilde{q}_L}^2-m_{\tilde{g}}^2)
\dot{B} (m_{\tilde{g}}^2, m_q^2,m_{\tilde{q}_L}^2) 
+2 m_{\tilde{g}}^2 (m_{\tilde{q}_R}^2-m_{\tilde{g}}^2)
\dot{B} (m_{\tilde{g}}^2, m_q^2,m_{\tilde{q}_R}^2) \}
 \nonumber \\
&& 
+( m_{\tilde{t}_1}^2-m_t^2) B_0(0, m_t^2,m_{\tilde{t}_1}^2)
+( m_{\tilde{t}_2}^2-m_t^2)  B_0(0, m_t^2,m_{\tilde{t}_2}^2) 
\nonumber \\
&&-(m_{\tilde{g}}^2+m_{\tilde{t}_1}^2-m_t^2) 
B_0(m_{\tilde{g}}^2, m_t^2,m_{\tilde{t}_1}^2)
-(m_{\tilde{g}}^2+m_{\tilde{t}_2}^2-
m_t^2) B_0(m_{\tilde{g}}^2, m_t^2,m_{\tilde{t}_2}^2)
\nonumber \\
&& +2 m_{\tilde{g}}^2 (m_{\tilde{t}_1}^2-m_{\tilde{g}}^2-
m_t^2+ 2 m_{\tilde{g}} m_t \sin(2 \theta) )
\dot{B} (m_{\tilde{g}}^2, m_t^2,m_{\tilde{t}_1}^2) 
\nonumber \\
&&+2 m_{\tilde{g}}^2 (m_{\tilde{t}_2}^2-m_{\tilde{g}}^2-m_t^2
-2 m_{\tilde{g}} m_t \sin(2 \theta) )
\dot{B} (m_{\tilde{g}}^2, m_t^2,m_{\tilde{t}_2}^2)\nonumber \\
&&\left. 
-6 m_{\tilde{g}}^2 B_0(0,m_{\tilde{g}}^2,\lambda^2)+
24 m_{\tilde{g}}^4 \dot{B}(m_{\tilde{g}}^2,m_{\tilde{g}}^2,
\lambda^2) \right), \nonumber \\
Z_R^{\tilde{g}}&=&Z_L^{\tilde{g}}, \nonumber \\
Z_{\tilde{t}_1}&=&{ 1\over 6 \pi^2} \left(
-B_0(m_{\tilde{t}_1}^2,m_{\tilde{g}}^2,m_t^2)+
B_0(m_{\tilde{t}_1}^2,\lambda^2,m_{\tilde{t}_1}^2) \right .\nonumber \\
&&+(m_{\tilde{g}}^2-m_{\tilde{t}_1}^2 +m_t^2) 
\dot{B} (m_{\tilde{t}_1}^2,m_{\tilde{g}}^2,m_t^2)
+2 m_{\tilde{t}_1}^2 \dot{B} (m_{\tilde{t}_1}^2, \lambda^2, m_{\tilde{t}_1}^2) \nonumber \\
&&\left.  -2 m_{\tilde{g}} m_t \sin(2 \theta) \dot{B} (m_{\tilde{t}_1}^2,m_{\tilde{g}}^2, m_t^2) \right).
\end{eqnarray}
We found the on mass shell definition for the counterterm of stops mixing,
\begin{eqnarray}
\delta \theta &=&
{1\over 2}(\delta Z^{12}+\delta Z^{21})={1\over 2} 
{\sum^{\tilde{t}_1\tilde{t}_2}(m_{\tilde{t}_2}^2)-
\sum^{\tilde{t}_1\tilde{t}_2}(m_{\tilde{t}_1}^2)
\over m_{\tilde{t}_2}^2-m_{\tilde{t}_1}^2} \nonumber \\
&=&{m_t m_{\tilde{g}} \cos(2 \theta) \over 6 \pi^2 (m_{\tilde{t}_2}^2-m_{\tilde{t}_1}^2)}
\left( (B_0(m_{\tilde{t}_2}^2,m_{\tilde{g}}^2,m_t^2) -
B_0(m_{\tilde{t}_1}^2,m_{\tilde{g}}^2,m_t^2) \right)
\end{eqnarray}
is appropriate for our calculation, which has been adopted in\cite{mix}. 
One can exam that such a $\delta \theta $ is UV convergent, so that it is renormalization group ( scale ) independent at one loop level.
As to the counterterm of the Yukawa coupling ${g^y_s}$ of ${SU(3)}_c$, we prefer to the treatment in \cite{Beenakker},
\begin{equation}
\delta g^y_s =-\frac{\alpha_s(\mu_R)}{4\pi} \{ [ \Delta - \log (\frac{\mu_R^2}{\mu^2}) ] \frac{\beta^{susy}_0}{2} - (\frac{2}{3} N_c - \frac{C_F}{2} ) \}. 
\end{equation}
However there are fewer virtual particles decoupled in our scenario with light 
gluino and light squarks,
\begin{equation}
\Delta g^y_s =\frac{\alpha_s(\mu_R)}{4\pi} 
[
 \frac{1}{3} \log \frac{\mu_R^2}{m^2_t}
+ \frac{1}{12} \log \frac{\mu_R^2}{m^2_{\tilde t_1}}
+ \frac{1}{12} \log \frac{\mu_R^2}{m^2_{\tilde t_2}}
 + \frac{n_f -1 }{6} \log \frac{\mu_R^2}{m^2_{\tilde q}} ]
\end{equation}
 The $f^d_i$ in Eq. \ref{total1} 
are the contributions from the 3-points diagrams,
\begin{eqnarray}
f_1^d &=& {\cos\theta-\sin\theta \over 96 \sqrt{2} \pi^2} \left(
(9-\sin(2 \theta)) B_0(m_t^2,m_{\tilde{g}}^2,m_{\tilde{t}_1}^2)+
\sin(2 \theta) B_0(m_t^2,m_{\tilde{g}}^2,m_{\tilde{t}_2}^2) \right. \nonumber \\
&&+35 B_0(m_t^2,\lambda^2,m_t^2)+ 18 m_{\tilde{g}}(3 m_{\tilde{g}}+m_t)C_0^{(1)}
-18 (m_t^2-m_{\tilde{t}_1}^2-m_{\tilde{g}}^2) C_0^{(2)}
\nonumber \\
&&+(-m_t^2-m_{\tilde{t}_1}^2+m_{\tilde{g}}^2) C_0^{(3)}
-2 m_t^2 \sin(2 \theta) C_0^{(4)}+ 2  m_t (-m_{\tilde{g}}+m_t
+m_t \sin(2 \theta)) C_0^{(5)} \nonumber \\
&&+18 (-2 k_1.p+3 m_{\tilde{g}}^2+m_{\tilde{g}} m_t) C_1^{(1)}
+18 (-2 k_1.p+2 m_{\tilde{g}}^2+m_{\tilde{g}} m_t) C_1^{(2)}
-2 m_{\tilde{g}} m_t C_1^{(3)} \nonumber \\
&&-2 m_{\tilde{g}}( m_{\tilde{g}}+(m_{\tilde{g}}-m_t) \sin(2 \theta)) C_1^{(4)}
-2  m_{\tilde{g}}( m_t +(m_t-m_{\tilde{g}})\sin(2 \theta)) C_1^{(5)} \nonumber \\
&&
+18 (-4 k_1.p+3 m_{\tilde{g}}^2+ m_t^2) C_2^{(1)}
+18 (-3 k_1.p+2 m_{\tilde{g}}^2+m_t^2) C_2^{(2)}
-2 (k_1.p-  m_t^2) C_2^{(3)} \nonumber \\
&&
-2 (m_{\tilde{g}}^2 - m_{\tilde{g}} m_t+ (-k_1.p +
m_{\tilde{g}}^2-m_{\tilde{g}} m_t+m_t^2) \sin(2 \theta)) C_2^{(4)}\nonumber \\
&&\left.-2 (-m_t^2 +m_{\tilde{g}} m_t+ (k_1.p -
m_{\tilde{g}}^2+m_{\tilde{g}} m_t-m_t^2) \sin(2 \theta)) C_2^{(5)} 
\right), \nonumber \\
f_2^d &=& {\cos\theta+\sin\theta \over 96 \sqrt{2} \pi^2} \left(
-(9+\sin(2 \theta)) B_0(m_t^2,m_{\tilde{g}}^2,m_{\tilde{t}_1}^2)+
\sin(2 \theta) B_0(m_t^2,m_{\tilde{g}}^2,m_{\tilde{t}_2}^2) \right. \nonumber \\
&&-35 B_0(m_t^2,\lambda^2,m_t^2)+ 18 m_{\tilde{g}}(-3 m_{\tilde{g}}+m_t)C_0^{(1)}
+18 (m_t^2-m_{\tilde{t}_1}^2-m_{\tilde{g}}^2) C_0^{(2)}
\nonumber \\
&&-(-m_t^2-m_{\tilde{t}_1}^2+m_{\tilde{g}}^2) C_0^{(3)}
-2 m_t^2 \sin(2 \theta) C_0^{(4)}+ 2  m_t (-m_{\tilde{g}}-m_t
+m_t \sin(2 \theta)) C_0^{(5)} \nonumber \\
&&+18 (2 k_1.p-3 m_{\tilde{g}}^2+m_{\tilde{g}} m_t) C_1^{(1)}
+18 (2 k_1.p-2 m_{\tilde{g}}^2+m_{\tilde{g}} m_t) C_1^{(2)}
-2 m_{\tilde{g}} m_t C_1^{(3)} \nonumber \\
&&+2 m_{\tilde{g}}( m_{\tilde{g}}-(m_{\tilde{g}}+m_t) \sin(2 \theta)) C_1^{(4)}
+2  m_{\tilde{g}}( -m_t +(m_t+m_{\tilde{g}})\sin(2 \theta)) C_1^{(5)} \nonumber \\
&&
-18 (-4 k_1.p+3 m_{\tilde{g}}^2+ m_t^2) C_2^{(1)}
-18 (-3 k_1.p+2 m_{\tilde{g}}^2+m_t^2) C_2^{(2)}
+2 (k_1.p-  m_t^2) C_2^{(3)} \nonumber \\
&&
+2 (m_{\tilde{g}}^2 + m_{\tilde{g}} m_t+ (k_1.p -
m_{\tilde{g}}^2-m_{\tilde{g}} m_t - m_t^2) \sin(2 \theta)) C_2^{(4)}\nonumber \\
&&\left.+2 (-m_t^2 -m_{\tilde{g}} m_t+ (-k_1.p +
m_{\tilde{g}}^2+m_{\tilde{g}} m_t+m_t^2) \sin(2 \theta)) C_2^{(5)} 
\right).
\end{eqnarray}
In the presentation of the formulas above,
we have used notations, 
\begin{eqnarray}
\Delta&=&{2 \over 4-D} -\gamma_E+ln(4 \pi),  \nonumber \\
\beta^{susy}_0&=&3 N_c - n_f = \beta^{light} + \beta^{heavy} \nonumber \\
k_1.p&=& { m_t^2-m_{\tilde{t}_1}^2+m_{\tilde{g}}^2 \over 2},  \nonumber \\
\dot B(p^2,m_a^2,m_b^2)&=&\partial B_0(p^2,m_a^2, m_b^2)/\partial p^2,
\end{eqnarray}
$\beta^{heavy} $ is for the decoupled virtual particles, which will not contribute to the scale evolution of $\alpha_s$. The definitions of the scalar integrals $Bs$ and $Cs$ could be 
found in Ref. \cite{r13,denner}, 
and the indexes $(1)-(5)$ of C functions have the variables
$$( m_{\tilde{g}}^2, m_t^2, m_{\tilde{t}_1}^2, m_{\tilde{g}}^2, 
\lambda^2,m_t^2),
( m_{\tilde{g}}^2, m_t^2, m_{\tilde{t}_1}^2, \lambda^2, 
m_{\tilde{g}}^2, m_{\tilde{t}_1}^2),
( m_{\tilde{g}}^2, m_t^2, m_{\tilde{t}_1}^2, m_{\tilde{t}_1}^2, 
m_t^2, \lambda^2),$$
$$( m_{\tilde{g}}^2, m_t^2, m_{\tilde{t}_1}^2,
m_t^2, m_{\tilde{t}_1}^2,m_{\tilde{g}}^2),
( m_{\tilde{g}}^2, m_t^2, m_{\tilde{t}_1}^2, m_t^2,m_{\tilde{t}_2}^2,
m_{\tilde{g}}^2),$$
respectively.

\subsection{Real corrections}
The infrared divergence arise from the virtual massless gluon corrections 
are compensated by
the real gluon bremsstrahlung corrections, i.e. the three-body decay
\begin{eqnarray}
t(p) \rightarrow \tilde{g}(k_1) \tilde{t}_1(k_2) g(k).
\end{eqnarray}   
The corresponding matrix element as given by the Feynman diagrams (Fig. 1 (f) ) is
\begin{eqnarray}
M_b&=&4 \sqrt{2} \pi \alpha_s  \bar{U}(k1) \{
-i f_{abc} T_c {1 \over 2 k.k_1} \rlap/\epsilon (\rlap/k+\rlap/k_1+m_{\tilde{g}})
(\sin\theta P_R-\cos\theta P_L) \nonumber \\
&&+ T_b T_a {1 \over -2 p.k}(\sin\theta P_R-\cos\theta P_L)
 (\rlap/k_1+\rlap/k_2+m_t) \rlap/\epsilon \nonumber \\
&&+T_a T_b {1 \over 2 k_2.k} (\sin\theta P_R-\cos\theta P_L)
(2 k_2+k).\epsilon \} U(p),
\end{eqnarray}   
where $\epsilon$ denotes the polarization vector of gluon. Squaring
the matrix element, performing
the polarization and color sum over the square of the amplitude and  
integrating over the phase space yields the complete
bremsstrahlung cross section as
\begin{eqnarray}
\Gamma_b &=& {\alpha_s^2 \over 24 \pi m_t} \{
48 \left[ 2 m_{\tilde{g}}^2 (m_{\tilde{g}}^2 -m_{\tilde{t}_1}^2
+m_t^2-2 m_{\tilde{g}} m_t \sin(2\theta))I_{11}
+2 m_{\tilde{g}}^2 I_1+I^0_1 \right.
\nonumber \\
&&
\left.
-2 m_{\tilde{g}} m_t \sin(2\theta)I_{1}\right]+
{64 \over 3} \left[ 2 m_t^2 (-m_{\tilde{g}}^2 +m_{\tilde{t}_1}^2
-m_t^2+2 m_{\tilde{g}} m_t \sin(2\theta))I_{00} \right.
\nonumber \\
&&\left.  -2 m_t^2 I_0 - I^1_0+
2 m_{\tilde{g}} m_t \sin(2\theta)I_{0}\right] 
+{64 \over 3} \left[ 2 m_{\tilde{t}_1}^2  (-m_{\tilde{g}}^2 +m_{\tilde{t}_1}^2
-m_t^2 \right. \nonumber \\
&&+2 m_{\tilde{g}} m_t \sin(2\theta))I_{22} 
\left. +(-m_{\tilde{g}}^2 +3 m_{\tilde{t}_1}^2 -m_t^2+
2 m_{\tilde{g}} m_t \sin(2\theta)) I_2+ I \right]
\nonumber \\
&&+ {8 \over 3} \left[
-(2 m_{\tilde{g}}^4-4 m_{\tilde{g}}^2 m_{\tilde{t}_1}^2
+2  m_{\tilde{t}_1}^4- 2 m_t^4 -
4 m_{\tilde{g}} m_t\sin(2\theta) ( m_{\tilde{g}}^2
-m_{\tilde{t}_1}^2-m_t^2)) I_{02} \right. \nonumber \\
&& \left.
+( m_{\tilde{g}}^2
+m_{\tilde{t}_1}^2+m_t^2) I_2 +2 (m_{\tilde{t}_1}^2-m_{\tilde{g}}^2) I_0
+I -2 m_{\tilde{g}} m_t \sin(2\theta) (I_2-I_0) \right] \nonumber \\
&&+ 24 \left[
-2 ( m_{\tilde{g}}^2
-m_{\tilde{t}_1}^2+m_t^2)^2- 2m_{\tilde{g}} m_t\sin(2\theta) 
(m_{\tilde{g}}^2
-m_{\tilde{t}_1}^2+m_t^2)) I_{01} \right. \nonumber \\
&&\left. +2 (m_t^2-m_{\tilde{g}}^2)I_0 +
2 (m_{\tilde{t}_1}^2-m_t^2+ m_{\tilde{g}} m_t \sin(2\theta)
)(I_1+I_0)-2 I \right] \nonumber \\
&&+ 24 \left[
2(-m_{\tilde{g}}^4+(m_{\tilde{t}_1}^4-2 m_{\tilde{t}_1}^2
m_t^2+m_t^4+2 m_{\tilde{g}} m_t\sin(2\theta) ( m_{\tilde{g}}^2
+m_{\tilde{t}_1}^2-m_t^2)) I_{12} \right. \nonumber \\
&& \left. -
(m_{\tilde{g}}^2
+m_{\tilde{t}_1}^2+m_t^2) I_2
+2 (m_{\tilde{t}_1}^2-m_t^2)I_1 -I \right].
\end{eqnarray}   
Here the hard gluon has been included and the definition of function 
$I^{AB...}_{ab...}$ are
\begin{eqnarray}
I^{AB...}_{ab...}
= {1\over \pi^2} \int 
{d^3k_1 \over 2 k_1^0}
{d^3k_2 \over 2 k_2^0}
{d^3k \over 2 k^0}
\delta^4(p-k_1-k_2-k)
{(\pm k.p_A) (\pm k.P_B)...
\over (\pm k.p_a) (\pm k.P_b)...},
\end{eqnarray}
where the minus corresponds top quark and
plus corresponds gluino and stop\footnote{
It should be noticed that (D.11) and (D.12) in Ref. \cite{r13} must
be corrected as following: the first term in the square bracket of (D.11) 
must be replaced by the corresponding term in (D.12),
and {\em vice versa}}.

\section{Numerical Results} 
In the MSSM the mass eigenstates $\tilde{q}_1$ and $\tilde{q}_2$ of the 
squarks are related to the interaction eigenstates $\tilde{q}_L$ and $\tilde{q}_R$ 
by \cite{MSSM}
\begin{eqnarray}
\left(\begin{array}{c}
\tilde{q}_1 \\ \tilde{q}_2\end{array}\right)=
R^{\tilde{q}}\left(\begin{array}{c}
\tilde{q}_L \\ \tilde{q}_R\end{array}\right)\ \ \ \ \mbox{with}\ \ \ \
R^{\tilde{q}}=\left(\begin{array}{cc}
                   \cos\theta_{\tilde{q}} & \sin\theta_{\tilde{q}}\\
                   -\sin\theta_{\tilde{q}} & \cos\theta_{\tilde{q}}
                   \end{array}
                   \right).
\label{eq1}
\end{eqnarray}
Following the notation of \cite{MSSM}, the mixing angle $\theta_{\tilde{q}}$ 
and the masses $m_{\tilde{q}_{1,2}}$ can be calculated by diagonizing the following mass
matrices
\begin{eqnarray}
M^2_{\tilde{q}}=\left(\begin{array}{cc}
          M_{LL}^2 & M_{LR}^2\\
           M_{RL}^2 & M_{RR}^2
           \end{array} \right), \nonumber \\
M_{LL}^2=m_{\tilde{Q}}^2+m_q^2+m_{z}^2\cos 2\beta (I_q^{3L}-e_q\sin^2\theta_w),
\nonumber \\
M_{RR}^2= m_{\tilde{U},\tilde{D}}^2 +m_q^2+m_{z}^2\cos 2\beta e_q\sin^2\theta_w.
\label{eq2}
\end{eqnarray}
>From Eqs. \ref{eq1} and \ref{eq2}, $m_{\tilde{t}_{1,2}}$ and $\theta$ can be derived as
\begin{eqnarray}
m^2_{\tilde{t}_{1,2}}&=&{1\over 2}\left[ M^2_{LL}+
M^2_{RR}\mp \sqrt{
(M^2_{LL}-M^2_{RR})^2+4 M^4_{LR}}\right] \nonumber \\
\tan\theta&=&{m^2_{\tilde{t}_1}-M^2_{LL} \over M^2_{LR}}.
\label{eq3}
\end{eqnarray}

For the numerical calculations, the renormalization scale is set at 
the mass of decay particle, $\mu_R =m_t=176.0$ GeV  $\alpha_S=0.108$.
We also evaluated at $\mu_R = m_t /2 \alpha_s = .12$ and found the scale dependence very weak. The soft SUSY breaking mass terms are always naturally assumed as $m_{\tilde{U}} = m_{\tilde
{D}}=m_{\tilde{Q}}=m_S$. For simplicity, we also define a variable 
$r= M_{LL}^2/M_{LR}^2$.

In Fig. \ref{fig2}, we show the branching ratios and the QCD
relative corrections as function of $m_{\tilde{t}_1}$,
assuming $r=0.9$. The definition of branching ratio and 
relative correction are
$Br(t \rightarrow \tilde{g} \tilde{t}_1)
=\frac{1}{1+
\Gamma (t \rightarrow b W^+) / \Gamma (t \rightarrow \tilde{g} \tilde{t}_1) }$
and 
$\delta=( \Gamma_1 + \Gamma_b ) / \Gamma_0$, respectively.
>From these curves, one can see that the
branching ratios, which are 
not sensitive to the light gluino mass allowed by experiments, 
decrease with the increment of $m_{\tilde{t}_1}$,
which is the natural consequence of the shrink of the phase space.
One also can see that, the decay width could be large enough to compare with 
the $t \rightarrow b W^+$ channel in a large range of parameter space. 
On the other hand, the QCD corrections
always enhanced the decay widths, for 
$m_{\tilde{t}_1}= 160$ GeV, the corrections exceed $40 \%$.
   
In Fig. \ref{fig3}, we exam the result for its dependence upon the decoupled particles, by alternating the way for the mass of the lighter stop to vary.
In these figures, $r$ was scanned with fixed soft SUSY breaking mass term $m_S$ as $300$ GeV. 
We can see these results are very close to Fig. \ref{fig2}, which
is due to the fact that the contributions from other squarks (relatively
heavy) are less important.

In summary, we have calculated the decay width of 
the process $t \rightarrow 
\tilde{g} \tilde{t}_1$ including the one loop QCD corrections.

>From our numerical examples, we can see that the QCD corrections
could enhance the decay widths over $30 \%$ in a very
large mass range of the lighter stop, and the decay
widths of the process could be larger than that of the channel 
$t \rightarrow b W^+$.
As a supplement of stop pair production at LEP 2000 or Tevatron, such a single
 stop ( top ) channel might make the top phenomenology more rich if it is allowed 
kinetic ally.

\section*{Acknowledgments}
This work was supported in part by the post doctoral foundation
of China and S.H. Zhu gratefully acknowledges the
support of  K.C. Wong Education Foundation, Hong Kong.

\newpage

\epsfxsize= 7.8cm
\centerline{\epsffile{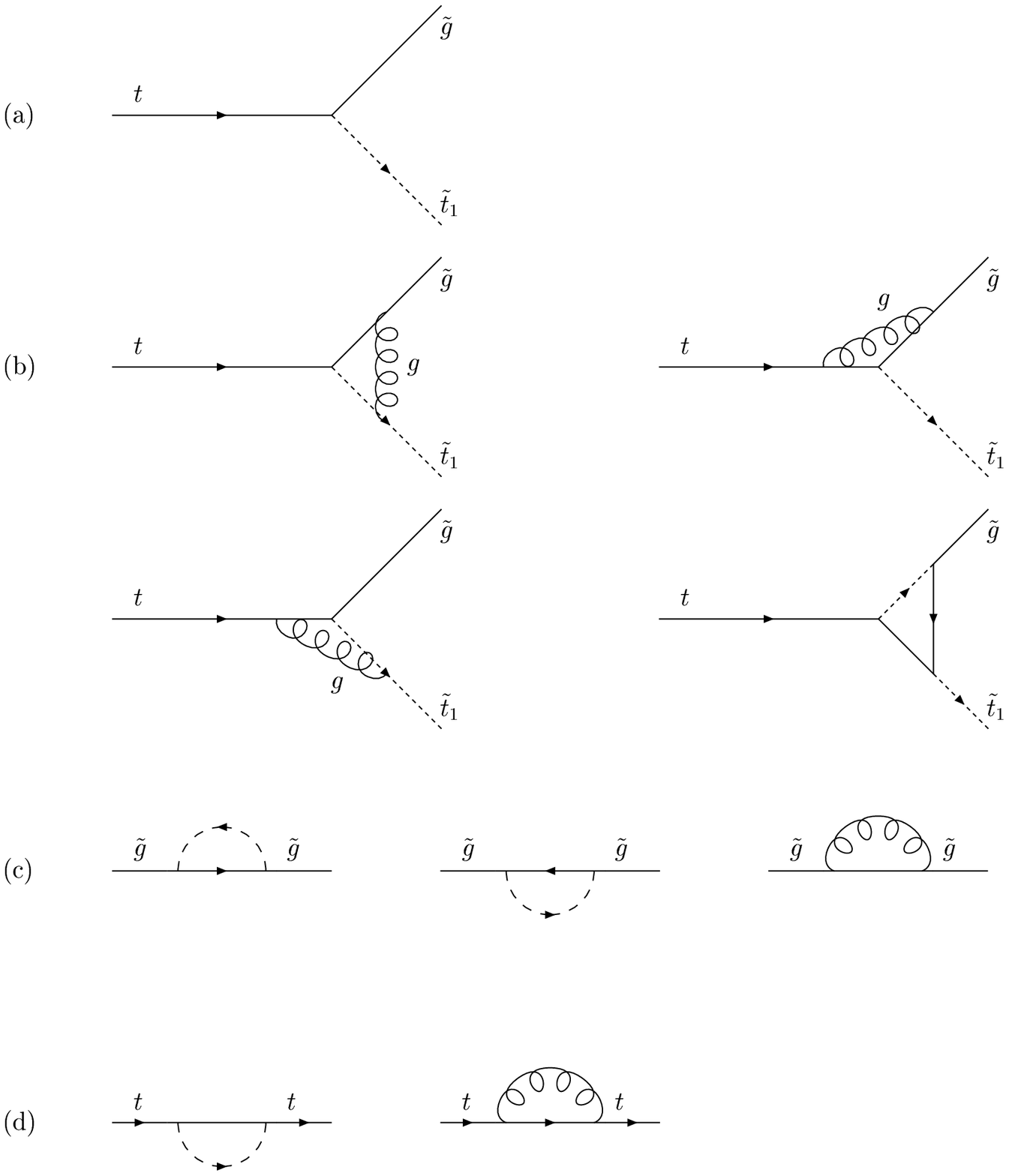}}
\label{feyn1}

\begin{figure}
\epsfxsize= 7.8cm
\centerline{\epsffile{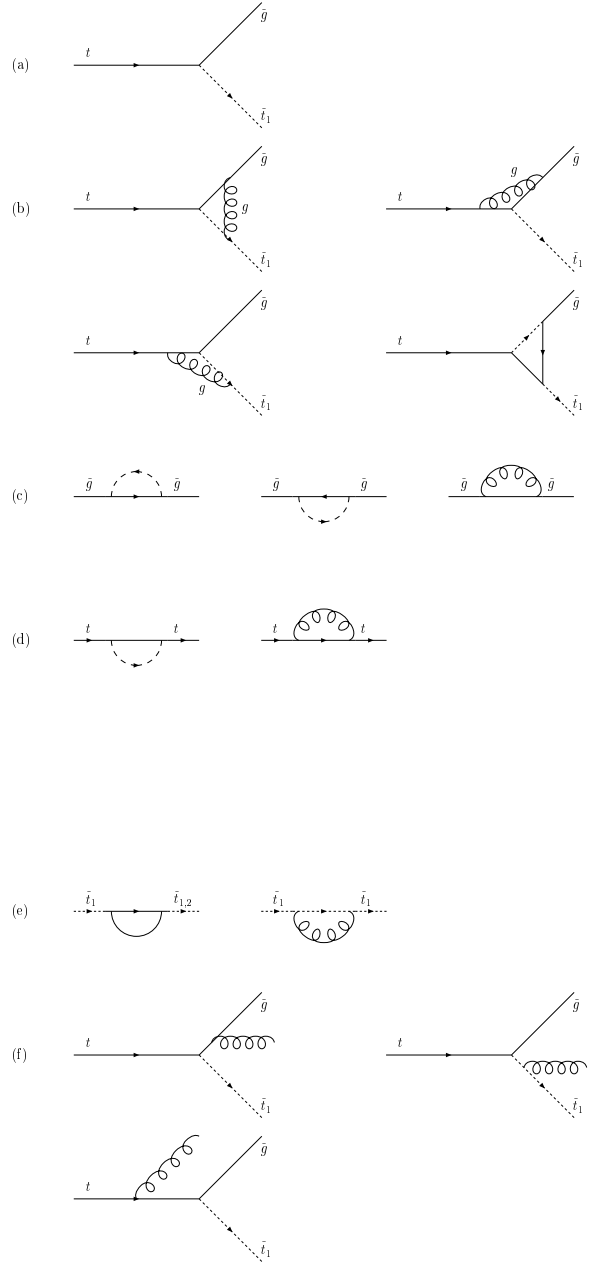}}
\caption[*]{Feynman diagrams for the process $t \rightarrow 
\tilde{g} \tilde{t}_1$.}
\label{feyn2}
\end{figure}

\begin{figure}
\epsfxsize= 18cm
\centerline{\epsffile{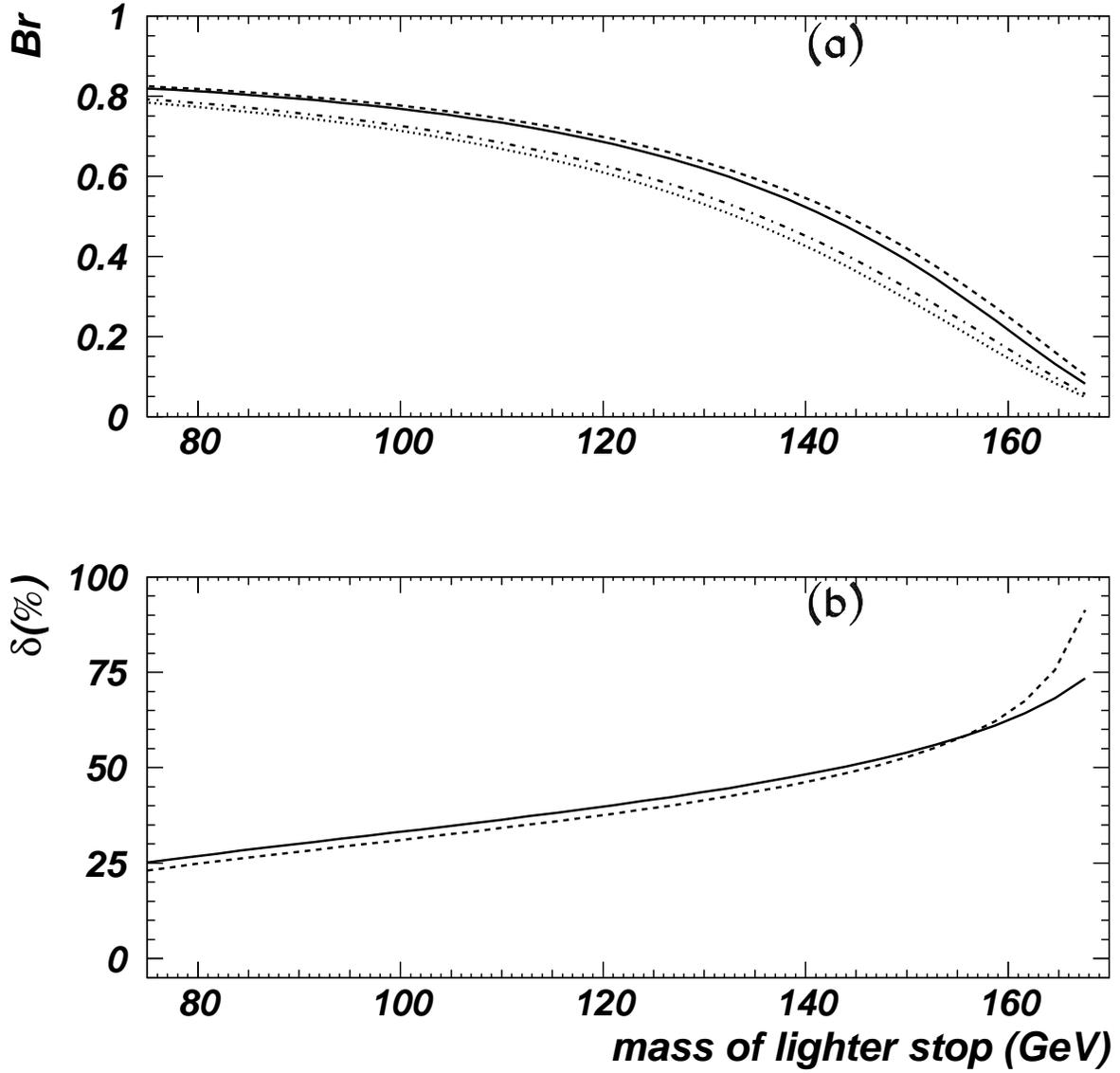}}
\caption[]{ Branching ratios and relative corrections 
as function of $m_{\tilde{t}_1}$, where $r=0.9$. {\bf (a)},
the solid ( dashed) line represents the decay widths including 
QCD corrections for $m_{\tilde{g}}= 1.0$ ($5.0$) GeV, and the dotted 
(dot-dashed) line denotes the tree level decay widths for $m_{\tilde{g}}= 1.0 $ ($5.0$) GeV;
{\bf (b)}, the solid (dashed) line represent the relative corrections
 for $m_{\tilde{g}}= 1.0$ ($5.0$) GeV }
\label{fig2}
\end{figure}

\begin{figure}
\epsfxsize= 18cm
\centerline{\epsffile{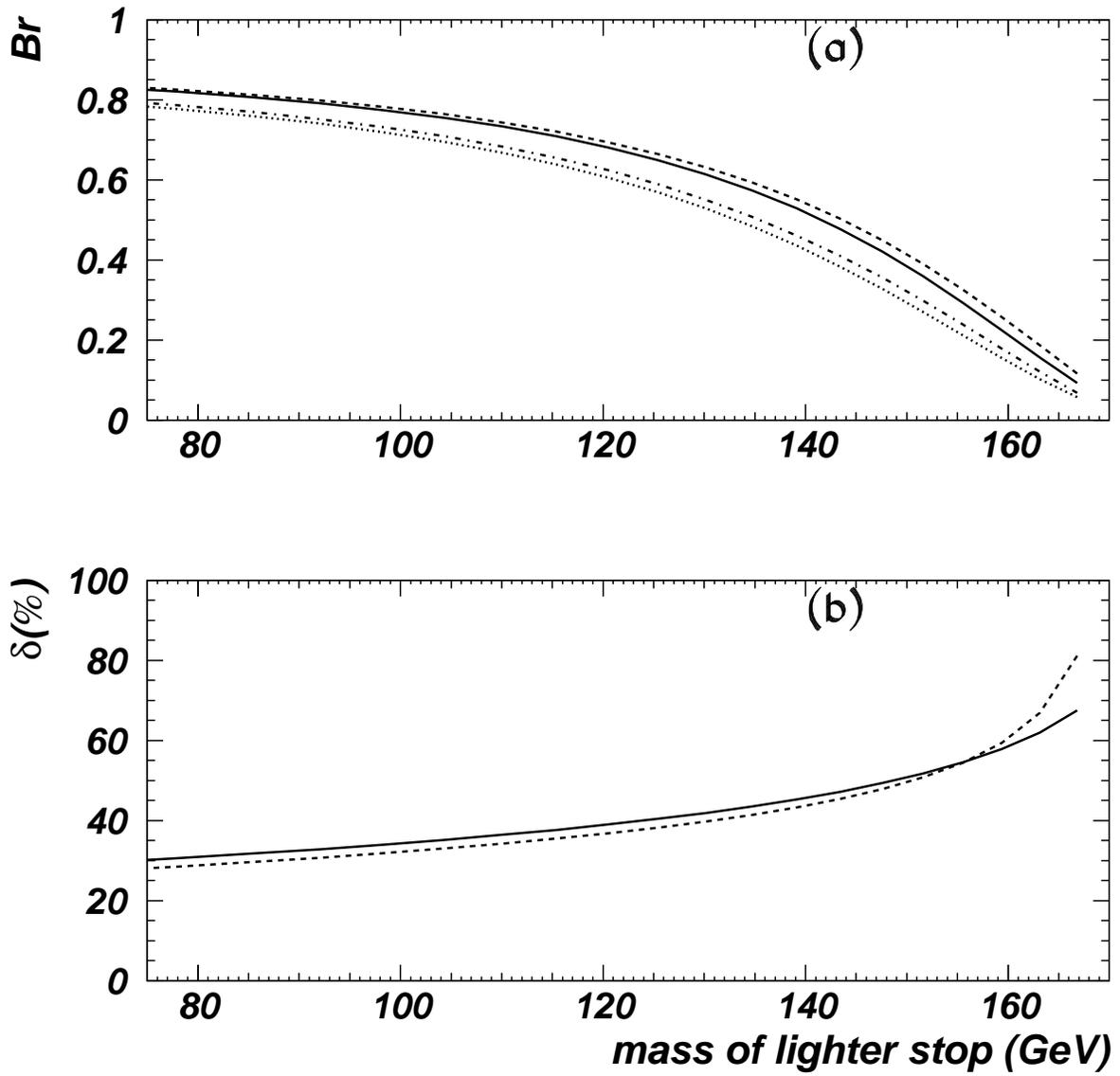}}
\caption[]{ see caption of  Fig. \ref{fig2} but fixed the soft SUSY
breaking mass term $m_S$ as $300$ GeV.
}
\label{fig3}
\end{figure}


\end{document}